\documentclass[a4paper,12pt]{article}
\usepackage{latexsym}
\usepackage{amsmath,amssymb,bm,ascmac,color,calc,subfigure}
\usepackage[mathscr]{eucal}
\usepackage{subfigure}
\usepackage{graphicx}
\usepackage{cite}
\usepackage{lineno}


\def\sla#1{\rlap{\kern .15em /}#1}

\begin{document}
\begin{titlepage}

\begin{flushright}
\begin{tabular}{l}
KEK-TH-1829
\end{tabular}
\end{flushright}

\vspace*{2cm}

\begin{center}
{\large\bf 
Evolution variable dependence of jet substructure
}\\[15mm]

Yasuhito Sakaki\footnote{E-mail: sakakiy@post.kek.jp}\\ \bigskip

{\em KEK Theory Center and Sokendai, Tsukuba, Ibaraki 305-0801, Japan}
\end{center}

\vspace{1cm}
\begin{abstract}
Studies on jet substructure have evolved significantly in recent years. Jet substructure is essentially determined by QCD radiations and non-perturbative effects. Predictions of jet substructure are usually different among Monte Carlo event generators, and 
are governed by the parton shower algorithm implemented. 
For leading logarithmic parton shower, even though one of the core variables is the evolution variable, its choice is not unique. 
We examine evolution variable dependence of the jet substructure by developing a parton shower generator that interpolates between different evolution variables using a parameter $\alpha$. Jet shape variables and associated jet rates for quark and gluon jets are used to demonstrate the $\alpha$-dependence of the jet substructure. We find angular ordered shower predicts wider jets, while relative transverse momentum ($p_{\bot}$) ordered shower predicts narrower jets. This is qualitatively in agreement with the missing phase space of $p_{\bot}$ ordered showers. Such difference can be reduced by tuning other parameters of the showering algorithm, especially in the low energy region, while the difference tends to increase for high energy jets.  
\end{abstract}
\end{titlepage}

\hrule
\tableofcontents
\vskip .2in
\hrule
\vskip .4in

\renewcommand\thefootnote{\roman{footnote}}
\section{Introduction}
The determination of different observables related to QCD jets is essential in studying the outcomes of high energy collision experiments. 
Successful predictions for such jet-variables have been achieved by using a combination of perturbative calculations at fixed order, parton shower algorithms, matrix-element and parton-shower matching algorithms and hadronization models. Study of jet substructure has also evolved significantly in recent times \cite{Abdesselam:2010pt, Altheimer:2012mn, Altheimer:2013yza, Adams:2015hiv}. Jet substructure techniques are particularly useful in identifying the origin of jet(s) in the hard process \cite{Seymour:1993mx, Butterworth:2002tt,Butterworth:2008iy,Thaler:2008ju,Kaplan:2008ie,Almeida:2008yp,Plehn:2009rk,Gallicchio:2010sw,Almeida:2010pa,Plehn:2010st,Thaler:2010tr,Soper:2011cr}, and also in removing contamination from pile-up or underlying event \cite{
Butterworth:2008iy,Cacciari:2007fd,Ellis:2009me,Krohn:2009th,Alon:2011xb,Soyez:2012hv,Larkoski:2014wba}.

The discrimination of quark-initiated jets from gluon-initiated ones is an important subject involving jet substructure, and has a lot of potential in improving the search for new physics. Different methods for quark-gluon tagging have been devised \cite{Gallicchio:2011xq,Gallicchio:2012ez,Krohn:2012fg,Chatrchyan:2012sn}, with corresponding performance studies \cite{CMS:2013kfa,CMS:2013wea,Aad:2014gea} for the Large Hadron Collider (LHC). Theoretical estimates for the performance of such tagging algorithms are primarily carried out with the help of Monte Carlo (MC) simulation tools, such as, {\tt Pythia} \cite{Sjostrand:2006za, Sjostrand:2007gs}, {\tt Herwig} \cite{Corcella:2000bw, Bahr:2008pv} and {\tt Sherpa} \cite{Gleisberg:2008ta}. Even though qualitative features are in agreement, differences in the predictions of the different MC's have been noted as far as quantitative estimates of the quark-gluon tagger performance is concerned. The primary reason for this can be traced back to the fact that the distribution of observables related to gluon jets varies significantly across the MC's, while those for the quark jet are largely similar. One possible cause of such a feature might be that while tuning the parameters of the MC generators, the precise jet data from the Large Electron-Positron Collider (LEP) have been crucial, and at leading order in electron-positron collision, the jet data is dominantly from quark-initiated processes. As far as the LEP data is concerned, the properly tuned versions of the MC's have been successful in achieving very good agreement with the jet data and are also consistent among each other, even in the soft-collinear and the non-perturbative regions.

Recent studies carried out by the ATLAS and CMS collaborations indicate that the data on certain observables related to quark-gluon tagging lies in between the predictions of the two MC generators {\tt Pythia} and {\tt Herwig} \cite{Aad:2014gea,ATLAS:2012am,Khachatryan:2014hpa}. Although it might be difficult to pinpoint the reason for such differences in the jet substructure observables predicted by different generators, understanding the difference between the central components of the MC's can be useful in developing more precise simulation tools. To this end, at a first order, if we postpone the consideration of the non-perturbative and underlying event effects for simplicity, the substructure of a quark or a gluon jet is governed by the pattern of QCD radiation, which is controlled by the parton shower algorithm. One of the core variables of a parton shower is the evolution variable, different choices for which are made in different MC's. In this study, our aim is to understand the effect of modifying the evolution variable and access its impact on jet substructure observables. We also ask the question whether certain choice of evolution variables can better reproduce the data on quark-gluon tagging observables, as discussed above.

With this goal in mind, we simulate jet substructure related observables with the following generalized evolution variable:
\begin{align}
Q_{\alpha}^2 = [4z(1-z)]^{\alpha} q^2, \label{a_def}
\end{align}
where, $\alpha$ is treated as a free parameter. For final state radiation, the above variable with $\alpha=1$ and $-1$ correspond to the evolution variables employed in {\tt Pythia8} and {\tt Herwig++} respectively. In Sec.~\ref{Formalism}, we provide further details on the framework used to implement this evolution variable in our parton shower program. In Sec.~\ref{Emission}, we show properties of QCD radiations generated by a given $Q_{\alpha}$, and discuss the correlation pattern between such radiation properties and the resulting behaviour of one important jet shape observable, $C_1^{(\beta)}$~\cite{Larkoski:2013eya}. In Sec.~\ref{Alpha}, we show $\alpha$-dependence of $C_1^{(\beta)}$ distributions and the associated jet rate observable~\cite{Bhattacherjee:2015psa} with tuned values of the parton shower parameters. We summarize our findings in Sec.~\ref{Conclusions}.

\section{Formalism}
\label{Formalism}

The evolution variable for the final state radiation of light partons used in our analysis is defined in Eq.~(\ref{a_def}), 
where $z$ is the momentum fraction of one of the daughter partons and $q^2$ is the virtuality of the mother parton. 
The daughter partons are taken to be on-shell. 
The variable $Q_{\alpha}$ is parametrized by a continuous parameter $\alpha$, 
and we take the range as $\alpha \in [-1,1]$ in this study. 
$Q_{\alpha}$ with $\alpha=1$ and $-1$ correspond to {\tt Pythia8}'s evolution variable (i.e., relative transverse momentum) and {\tt Herwig++}'s one respectively. 
QCD radiations are governed by the DGLAP equation \cite{Gribov:1972ri,Altarelli:1977zs}. 
When we use $Q_{\alpha}$ as a scale variable, 
the evolution equation takes on a equivalent form for each $\alpha$ due to the following relation, 
\begin{align}
\frac{dQ_{\alpha}^2}{Q_{\alpha}^2} dz
=
\frac{dq^2}{q^2} dz.
\end{align}
We implement the general evolution variable $Q_{\alpha}$ for arbitrary $\alpha$ in a parton shower program, 
and calculate jet substructure observables. 
Even though there are various recent parton shower formalisms, e.g., dipole shower in {\tt Pythia8} \cite{Sjostrand:2004ef} or dipole-antenna shower in {\tt Vincia} \cite{Giele:2007di,Giele:2011cb}, 
we use in this study a traditional formalism based on Refs.~\cite{Bahr:2008pv,Gieseke:2003rz}, which is used in {\tt Herwig++}. 
In the following subsection, we describe the modification to the formalism in Refs~\cite{Bahr:2008pv,Gieseke:2003rz} required to have a parton shower with arbitrary $\alpha$.

\subsection{Phase space}
\label{Phase}
Consider an emission where a mother parton $a$ branches off into light or massless partons $b$ and $c$ $(a\to b c)$. 
We give an effective mass $m_{\rm qg}$ to the daughter partons to avoid singularities in the splitting functions. 
Then, upper and lower values of the {\it energy} fraction of one daughter parton $z_E^+$ and $z_E^-$ are given by 
\begin{align}
z_E^{\pm} = \frac{1}{2} 
\left(
1 \pm 
\sqrt{1-\frac{q^2}{E_a^2}}
\sqrt{1-\frac{4m_{\rm qg}^2}{q^2}}
\right), 
\end{align}
where $q^2$ is the virtuality of $a$ when $b$ and $c$ are on-shell, 
and $E_a$ is the energy of $a$. 
This gives a condition for the allowed region on the energy fraction $z_E$ and $Q_{\alpha}$ as 
\begin{align}
\frac{Q_{\rm min}^2}{Q_{\alpha}^2} w^{\alpha} + 
\frac{Q_{\alpha}^2}{Q_{\rm max}^2} w^{-\alpha} \leq 
w + \frac{Q_{\rm min}^2}{Q_{\rm max}^2}, ~~~~~
w = 4z_E(1-z_E),   \label{PScon}
\end{align}
where $Q_{\rm max}$ and $Q_{\rm min}$ are the maximal and minimal values for $Q_{\alpha}$. 
These are independent of $\alpha$, and given as 
\begin{align}
Q_{\rm max} = E_a, ~~~~~~
Q_{\rm min} = 2 m_{\rm qg}.  \label{maxmin}
\end{align}
Here, $z$ describes not the energy fraction but the light-cone momentum fraction as in Refs.~\cite{Bahr:2008pv,Gieseke:2003rz}. 
However, we have explicitly checked that these are approximately the same. 
Hence we use Eq.~(\ref{PScon}) with a substitutions, $z_E \to z$ in the generation of $Q_{\alpha}$ and $z$. 
The energy of the partons are known at the end of all branchings. 
So, we set $Q_{\rm max}$ in Eq.~(\ref{PScon}) to the energy of the initial hard scattering process, i.e., $\sqrt{s}/2$ for the first branching, and 
calculate by taking $z$ as the energy fraction for subsequent branchings. 
These choices ensure the required relation $\bold{p}_{\bot}^2 = Q_1^2 - Q_{\rm min}^2 \geq 0$, 
where $\bold{p}_{\bot}$ is the spatial component of the {\it relative transverse momentum} for each branchings, as defined in Ref.~\cite{Bahr:2008pv,Gieseke:2003rz}.

\begin{figure}[!t]
\begin{center}
\includegraphics[width=9.0cm, bb=0 0 759 609]{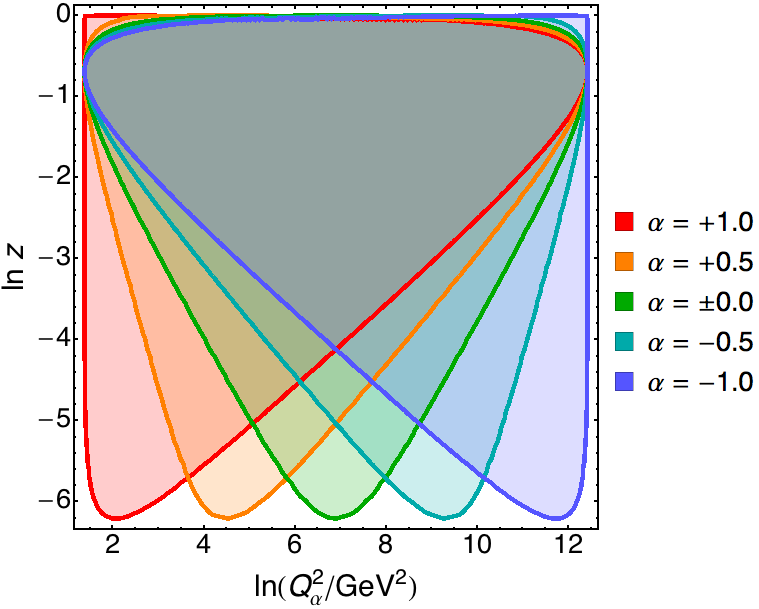}
\caption{{\footnotesize 
The allowed phase space in the $\ln z - \ln Q_{\alpha}$ plane for each choice of $\alpha$, with $E_a$ fixed at 500 GeV. 
}}
\label{ps}
\end{center}
\end{figure}
The allowed phase spaces in the $\ln z - \ln Q_{\alpha}$ plane for each choice of $\alpha$ is illustrated in Fig. \ref{ps}, where 
the parton energy $E_a$ is fixed at 500 GeV. 
At leading order, the parton branchings occur almost uniformly on this plane. 
The partons start from a high scale and evolve to low scale in timelike branchings, 
and the smaller $\alpha$ is, the larger the phase space becomes in the high scale region. 
So, when the evolution starts from a high scale, initial emissions tend to choose a high scale and soft emission for small $\alpha$.

\subsection{Starting scale}
We consider final states of either a light quark pair $(q\bar{q})$ or a gluon pair $(gg)$, with a center of mass energy of $\sqrt{s}$, 
and set the starting scale for the initial partons to their energy in the rest frame of the final state, i.e., $\sqrt{s}/2$. 
This is the maximal choice for the starting scale, see Eq.~(\ref{maxmin}).

Next, consider the sequential branchings $a \to bc$ and $b \to de$, with the scales of the branching given by $Q_{\alpha}$ and $Q_{\alpha,b}$ as;
\begin{align}
 Q_{\alpha}^2    &\simeq [4z(1-z)]^{\alpha} \times 2z(1-z) E_a^2 (1-\cos\theta_a), \\
 Q_{\alpha,b}^2 &\simeq [4z_b(1-z_b)]^{\alpha} \times 2z_b(1-z_b) E_b^2 (1-\cos\theta_b), 
\end{align}
where $\theta_a$ and $\theta_b$ are the angle between $b$ and $c$, and $d$ and $e$ respectively. 
The momentum fractions for the branchings $a \to bc$ and $b \to de$ are given by $z$ and $z_b$, and 
the energy of $a$ and $b$ are $E_a$ and $E_b \simeq z_b E_a$. 
By imposing the angular ordering $\theta_a > \theta_b $, we get 
\begin{align}
Q_{\alpha,b} 
&< 
Q_{\alpha} z
\left[
\frac
{4z(1-z)}
{4z_b(1-z_b)}
\right]^{-(\alpha+1)/2}, \\
& \leq 
Q_{\alpha} z 
[4z(1-z)]^{-(\alpha+1)/2}. \label{st1}
\end{align}
The right-hand side in Eq.~(\ref{st1}) can be greater than the previous scale $Q_{\alpha}$. 
To avoid this wrong of ordering the scale, we set the starting scale of the daughter parton $b$ as
\begin{align}
Q_{\alpha,b}^S = 
Q_{\alpha} 
{\rm min} (1, z [4z(1-z)]^{-(\alpha+1)/2}). 
\end{align}
The angular ordering is ensured by using this starting scale for $\alpha = -1$. 
However, angular ordered emission is not ensured for $\alpha \neq -1$. 
Such emissions are vetoed by hand as in {\tt Pythia6} \cite{Sjostrand:2006za}.

\subsection{Tunable parameters and other modifications}
\label{Tunable}
We use three parameters $\alpha_S(m_Z)$, $m_{\rm qg}$, and $r_{\rm cut}$ in our parton shower program. 
The first one is the strong coupling constant at the scale of the $Z$ boson mass. 
We use one loop running of $\alpha_S$ in our code. 
The argument of $\alpha_S$ is set to $p_{\bot} = 2^{-\alpha} [z(1-z)]^{(1-\alpha)/2} Q_{\alpha}$ thereby including the effects of subleading terms in the splitting functions. 
The value of $\alpha_S$ is significant to the predictions of jet substructure.
Larger values of $\alpha_S$ lead to high scale emissions, and jet shape distributions, e.g., the jet mass distribution shift to higher value regions. 
The value of $\alpha_S(m_Z)$ is set to 0.118 in {\tt Herwig++}, and
about $0.136 - 0.139$ for the final state radiation in {\tt Pyhtia8}. 
The second variable $m_{\rm qg}$ is the effective mass of the light partons and gluons to avoid soft-collinear singularities, 
which was introduced in Sec. \ref{Phase}. 
The third one is defined as
\begin{align}
r_{\rm cut} = \frac{Q_{\rm cut}}{Q_{\rm min}} = \frac{Q_{\rm cut}}{2m_{\rm qg}}, 
\end{align}
where $Q_{\rm cut}$ is a given scale where the evolution terminates.

We note in passing that, in our analysis, we neglect $g\to q\bar{q}$ branchings for simplicity, which affect distributions at the NLL order. 

\section{Emission property}
\label{Emission}
\begin{figure}[!t]
\begin{center}
\includegraphics[width=15.5cm, bb=0 0 670 605]{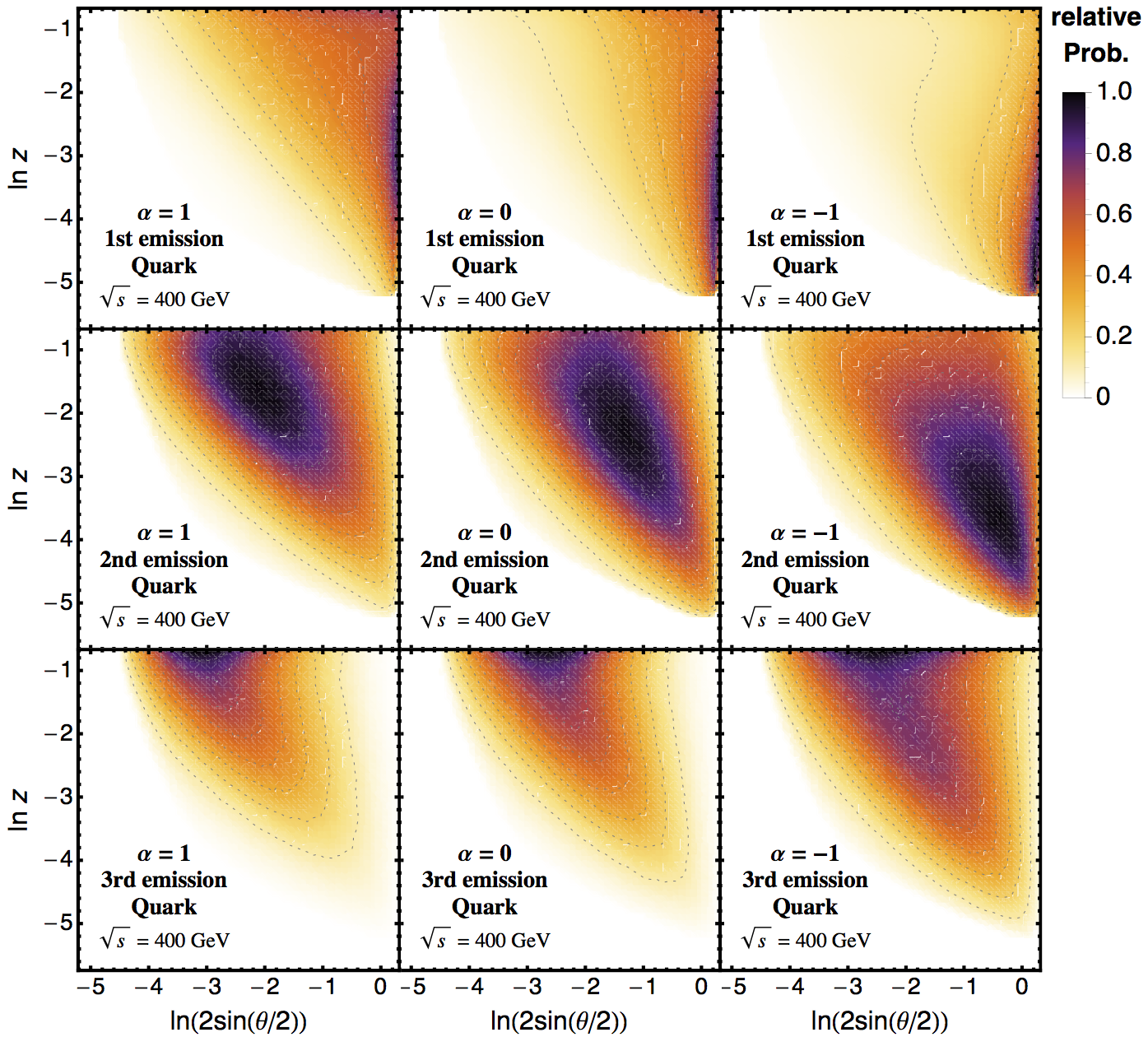}
\caption{{\footnotesize 
Emission probability in the $\ln z - \ln (2\sin(\theta/2))$ plane for quark jets. The top, center and bottom rows show the results for the first, second and third emissions, respectively. The second and third emission refer to the emissions from the harder parton produced by the first and second emissions respectively. 
}}
\label{emission_Q}
\end{center}
\end{figure}

\begin{figure}[!t]
\begin{center}
\includegraphics[width=15.5cm, bb=0 0 670 605]{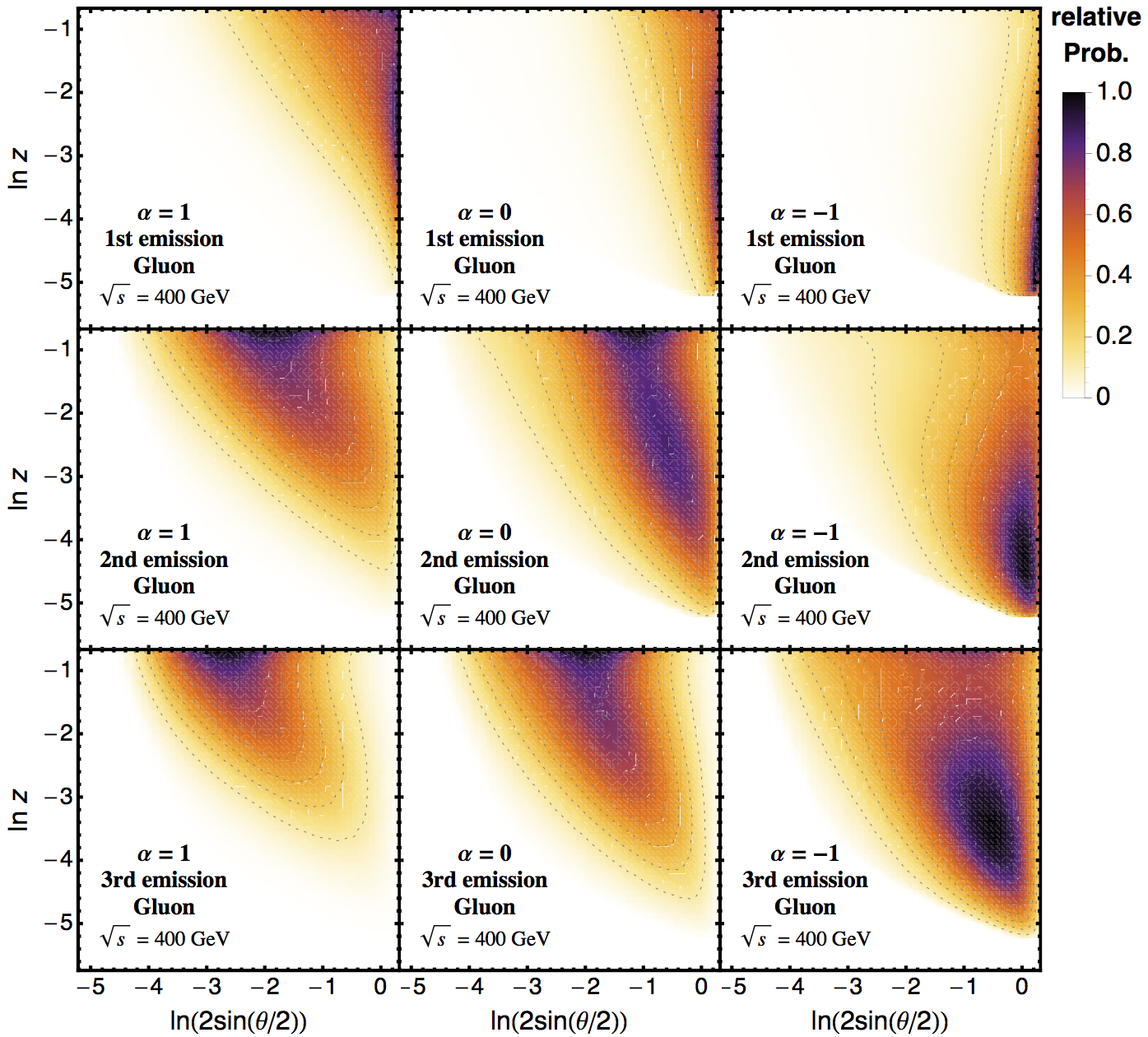}
\caption{{\footnotesize 
Same as Fig. \ref{emission_Q}, for gluon jets.
}}
\label{emission_G}
\end{center}
\end{figure}

Jet shape observables are important in examining the substructure of QCD jets. One of the recently studied jet shape observable is the two-point energy correlation function $C_1^{(\beta)}$ \cite{Larkoski:2013eya,Larkoski:2014uqa}, which can be defined in the rest frame of a parton pair as
\begin{align}
C_1^{(\beta)} = 
  \sum_{i<j \in {\rm jet}} 
  \frac{E_i E_j }{E_{\rm jet}^2}
\left(
  2\sin\frac{\theta_{ij}}{2}
\right)^{\beta}, 
\end{align}
where $E_i$ and $E_j$ are the energies of the particles labeled by $i$ and $j$ in the jet, $E_{\rm jet}$ is the jet energy, and $\theta_{ij}$ is the angle between $i$ and $j$. The sum runs over all distinct pairs of particles in the jet. The dominant contribution to this observable comes from the hardest emission in the jet, 
which is also the first emission in the jet \cite{Banfi:2004yd}. Neglecting all other emissions except for the hardest one, we get in the soft limit
\begin{align}
\ln C_1^{(\beta)} \simeq \ln z + \beta \ln \left(2\sin\frac{\theta}{2} \right),  \label{C1app}
\end{align}
where $z$ and $\theta$ are the smaller energy fraction and the angle of the hardest emission, respectively. Evidently from the above equation, studying the properties of the first emission in the jet on the $z - \theta$ plane will lead to an understanding of the behaviour of this jet shape.

Figs. \ref{emission_Q} and \ref{emission_G} show the emission probability on the $\ln z - \ln (2\sin(\theta/2))$ plane for quark and gluon jets respectively. The top, center and bottom rows show the results for the first, second and third emissions. Here, the second and third emissions refer to the emissions from the harder of the two partons produced by the first and second emissions respectively. 
We find that the equal-probability curves for the first emission plots are roughly given by the contours described by
\begin{align}
{\rm Const.} = \frac{\alpha + 1}{2} \ln z + \ln \left(2\sin\frac{\theta}{2} \right). 
\end{align}
This is because, the evolution variable, in other words, the {\it ordering} variable, in Eq.~(\ref{a_def}) can be written in the soft limit as
\begin{align}
\ln Q_{\alpha} = \frac{\alpha + 1}{2} \ln z + \ln \left(2\sin\frac{\theta}{2} \right) + {\rm Const.} 
\end{align}
It should be mentioned that the small $z$ regions are more favourable due to larger values of the strong coupling constant, $\alpha_S$. 
In the case of $\alpha=-1$, the evolution variable is given by $Q_{-1} \simeq E \times 2\sin(\theta/2)$, where $E$ is the energy of the mother parton. So, high scales also imply larger angles. As mentioned above, the emissions tend to prefer high scales and soft emissions for smaller values of $\alpha$. This is consistent with the results for the first emission with $\alpha=-1$ in Figs. \ref{emission_Q} and \ref{emission_G}.

Clearly, for the jet shape observable in question, we are mostly interested here in the first emission in a jet. When we set the jet radius to $R=0.4$, such emissions are distributed in the region described by $\ln(2\sin(\theta/2)) < -0.9$. The first emissions often fall outside a narrow jet, especially for small $\alpha$. Also, such emissions tend to be vetoed out in the parton shower-matrix element matching algorithms. Therefore, it is also important to look into the subsequent emissions. We find that the second and the third emissions also have a different distribution for each value of $\alpha$. This fact indicates that parton shower algorithms implementing different evolution variables would have different predictions for jet substructure. However, the results in Figs.~\ref{emission_Q} and \ref{emission_G} are obtained with the same set of inputs for the tunable parameters described in the previous section \footnote{
The distributions in Figs.~\ref{emission_Q} and \ref{emission_G} are obtained with $\alpha_S(m_Z)=0.12$, $m_{\rm qg}=1{\rm GeV}$, and $r_{\rm cut}=1$. 
} for all values of $\alpha$. In the next section, we employ a procedure to fit the values of these parameters for each $\alpha$ separately,  and show our results with the fitted values of the parton shower parameters for completeness.

\section{The $\alpha$ dependence}
\label{Alpha}
\subsection{Jet shape distribution}
\label{JetShape}

Jet shape distributions depend on the parameters $\alpha_S(m_Z)$, $m_{\rm qg}$, and $r_{\rm cut}$  introduced in Sec. \ref{Tunable}. These parameters are determined by performing a fit of the MC predictions to experimental data on several jet observables, for which the $e^+ e^- \to n$ jets data from LEP are particularly useful. Performing such a fit to the experimental data is, however, beyond the scope of the present study as this would require the implementation of a hadronization model in our parton shower code. Since the primary goal of this study is to examine between difference between parton shower algorithms using different evolution variables, as an alternative to real data, we utilize the $e^+e^- \to q\bar{q}$ events generated by {\tt Herwig++} with hadronization switched off as our {\it data} \footnote{To be specific, we use {\tt Herwig++ 2.7.1} with default tune, for the $u\bar{u}$ and $d\bar{d}$ parton level final states. }. 

The $C_1^{(0.5)}$, $C_1^{(2.0)}$ and $C_1^{(3.0)}$ distributions have been used to tune the above parameters. As mentioned in Sec.~\ref{Emission}, the first emission in the jet has a significant effect on the jet shape, which can be parametrized by the momentum fraction $z$ and the angle $\theta$. Therefore, two independent $C_1^{(\beta)}$ distributions contain the necessary information about the jet shapes. Here, we use three variables in order to further examine the $\beta$ dependence of the QCD jet substructure. 

Throughout this paper, jets are clustered using the generalized $k_t$ algorithm for $e^+e^-$ collisions using {\tt FastJet 3.1.1} \cite{Cacciari:2011ma}, the distance measure for which is defined as 
\begin{align}
d_{ij} = {\rm min}(E_i^{2p}, E_j^{2p})
\frac{1-\cos\theta_{ij}}{1-\cos R}, 
\end{align}
where $R$ is the jet radius parameter, and we use $p=-1$.

\begin{table}[t]
   \begin{center}
   \begin{tabular}{|c||c|c|c|}
      \hline
	$\alpha$	& $\alpha_S(m_Z)$ & $m_{\rm qg}${\footnotesize [GeV]}  & $r_{\rm cut}$ \\
      \hline \hline
      $+1.0$		& $0.132$	& $0.94$	& $1.00$	\\
      $+0.5$		& $0.126$	& $0.90$	& $1.00$	\\
      $\pm 0.0$		& $0.121$	& $0.84$	& $1.05$	\\
      $-0.5$			& $0.119$	& $0.83$	& $1.16$	\\
      $-1.0$			& $0.119$	& $0.85$	& $1.25$	\\
      \hline
   \end{tabular}
   \caption{\footnotesize 
Tuned values of the parton shower parameters for each choice of $\alpha$, obtained by fitting the $\ln C_1^{(0.5)}$, $\ln C_1^{(2.0)}$ and $\ln C_1^{(3.0)}$ distributions for quark jets with $R=0.4$  with an $e^+ e^-$ centre of mass energy of $\sqrt{s} = 200 {\rm GeV}$. The reference distributions are calculated by using $e^+e^- \to q\bar{q}$ events generated by {\tt Herwig++}}
   \label{FittedPara}
\end{center}
\end{table}

We firstly generate events using five choices for the evolution variable, $Q_{1}$, $Q_{0.5}$, $Q_{0}$, $Q_{-0.5}$ and $Q_{-1}$ at $\sqrt{s} = 200 {\rm GeV}$, where $\sqrt{s}$ denotes the center of mass energy in the $e^+e^-$ collisions. We calculate $\ln C_1^{(0.5)}$, $\ln C_1^{(2.0)}$ and  $\ln C_1^{(3.0)}$ distributions with $R=0.4$, and find the values of the parameters that minimize the $\chi^2$ variable computed using our results and the mock data generated by {\tt Herwig++}. Theoretical errors are assigned using a  flat distribution for each bin.  The best fit values of the parameters are shown in Table \ref{FittedPara}. We see that the larger $\alpha$ is, the larger the tuned value of $\alpha_S(m_Z)$ becomes. In other words, the {\tt Pythia8}-like case with $Q_{1}$ prefers a higher value of $\alpha_S(m_Z)$ compared to the {\tt Herwig}-like case with $Q_{-1}$. This qualitative behaviour is in agreement with the actual implementations found in {\tt Pythia8} and {\tt Herwig++}. It should be emphasized that the outcomes of this tuning procedure do not entirely reflect the Monte Carlo difference between {\tt Pythia8} and {\tt Herwig++}, as the the parton shower algorithm implemented in {\tt Pythia8} is different from ours.

\begin{figure}
\begin{center}
\includegraphics[width=14.5cm, bb=0 0 640 960]{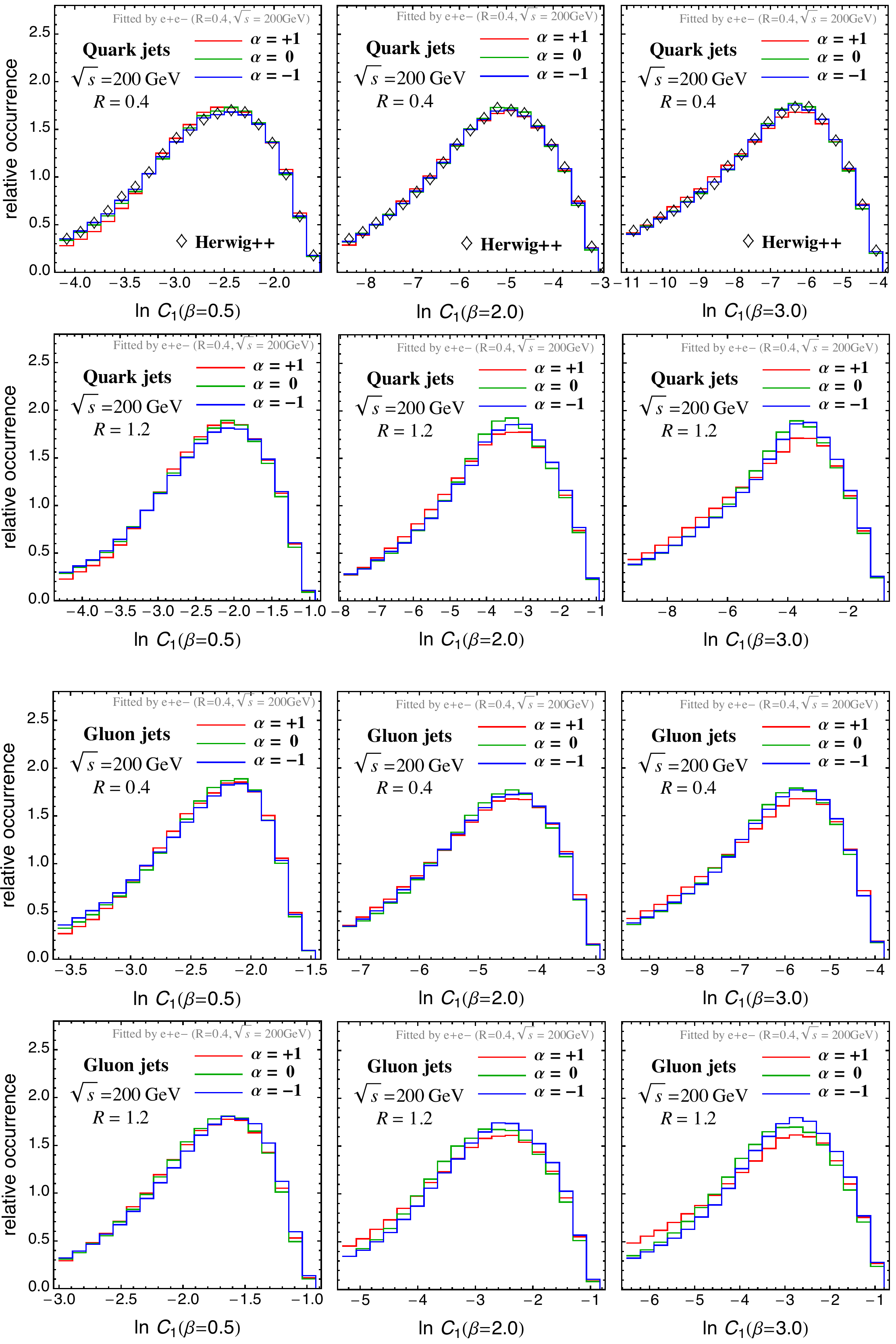}
\caption{{\footnotesize 
Distributions of $\ln C_1^{(0.5)}$, $\ln C_1^{(2.0)}$ and $\ln C_1^{(3.0)}$ for quark and gluon jets, with $R=0.4$ and $1.2$, at $\sqrt{s}=200$ GeV, as obtained using the parameter values shown in Table \ref{FittedPara}. 
}}
\label{dist200}
\end{center}
\end{figure}

\begin{figure}
\begin{center}
\includegraphics[width=14.5cm, bb=0 0 640 960]{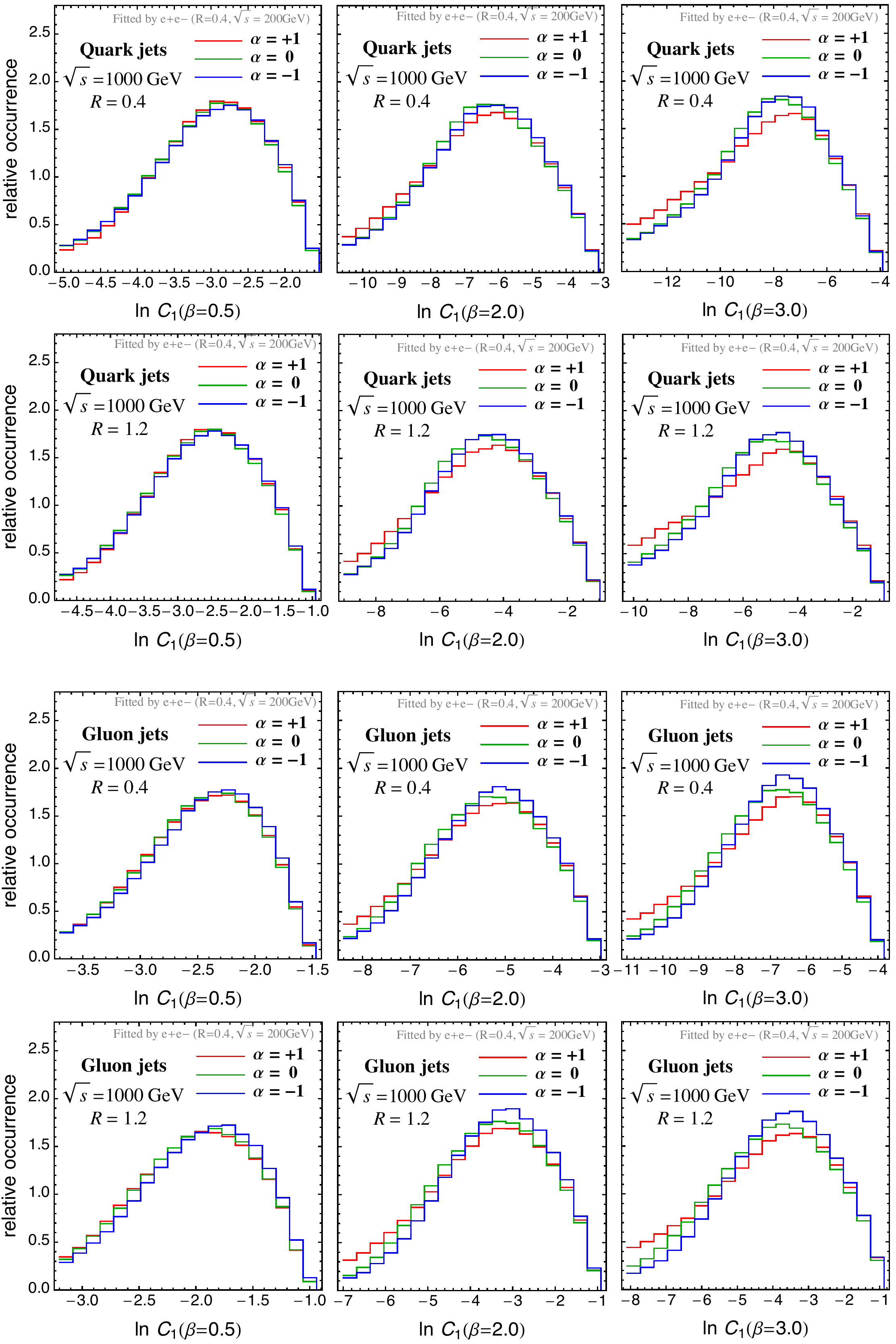}
\caption{{\footnotesize 
Same as Fig. \ref{dist200}, with a higher center of mass energy, $\sqrt{s}=1000$ GeV. 
}}
\label{dist1000}
\end{center}
\end{figure}

In Fig.~\ref{dist200}, the top row shows the fitted results, and hence the distributions are in good agreement with {\tt Herwig++} predictions. We also obtained the distributions for a fat jet (with $R=1.2$) and for gluon jets using the fitted values of the parameters shown in Table \ref{FittedPara}. For the same energy, the gluon jet distributions with $R=0.4$ are similar for each choice of the evolution variable. Small differences appear in the shapes predicted by different choices of $\alpha$ for the fat quark and gluon jets $(R=1.2)$. Fig.~\ref{dist1000} shows the same distributions as in Fig.~\ref{dist200}, with a higher value of the center of mass energy, $\sqrt{s}=1000$ GeV. As we can see, the $\alpha$-dependence of the shapes is found to be higher for higher energy jets.

\subsection{Wideness of soft emissions in jets}
The larger the parameter $\beta$ in $C_1^{(\beta)}$ is, the larger the differences become in Fig.~\ref{dist1000}. This implies that the wideness of the emissions, especially for the hardest emission in the jets, is different for each $\alpha$. This is because, the larger $\beta$ is, the larger the contribution to $C_1^{(\beta)}$ from the emission angle of the hardest emission becomes, which is understood from Eq.~(\ref{C1app}).

Associated jet rates defined in Ref.~\cite{Bhattacherjee:2015psa} directly reveal the wideness of the emissions in jets. Associated jets are jets nearby a hard jet, and are defined by two parameters, $R_a$ and $E_a$. Here, $R_a$ is the maximum allowed angle between the momentum directions of the hard jet and the associated jet, and $E_a$ is the minimum energy of the associated jets\footnote{
In Ref.~\cite{Bhattacherjee:2015psa}, for studies in hadron collisions, the parameter $p_a$ has been used to define associated jets instead of $E_a$, where $p_a$ is the minimum transverse momentum of the associated jets. However, for the $e^+e^-$ collisions studied in our paper, it is more suitable to use the energy variable.}. We set the value to $E_a = 20~{\rm GeV}$ in this study. 

A high probability for having no associated jet implies that the probability of wide emissions occurring around the hard jet is low. Such probabilities have been obtained by using {\tt Pythia8}, {\tt Pythia6}, and {\tt Herwig++}, and it has been found that the no associated jet probability predicted by {\tt Pythia} is higher than the one obtained with {\tt Herwig++} \cite{Bhattacherjee:2015psa}.

\begin{figure}[t]
\begin{center}
\includegraphics[width=6.7cm, bb=0 0 280 280]{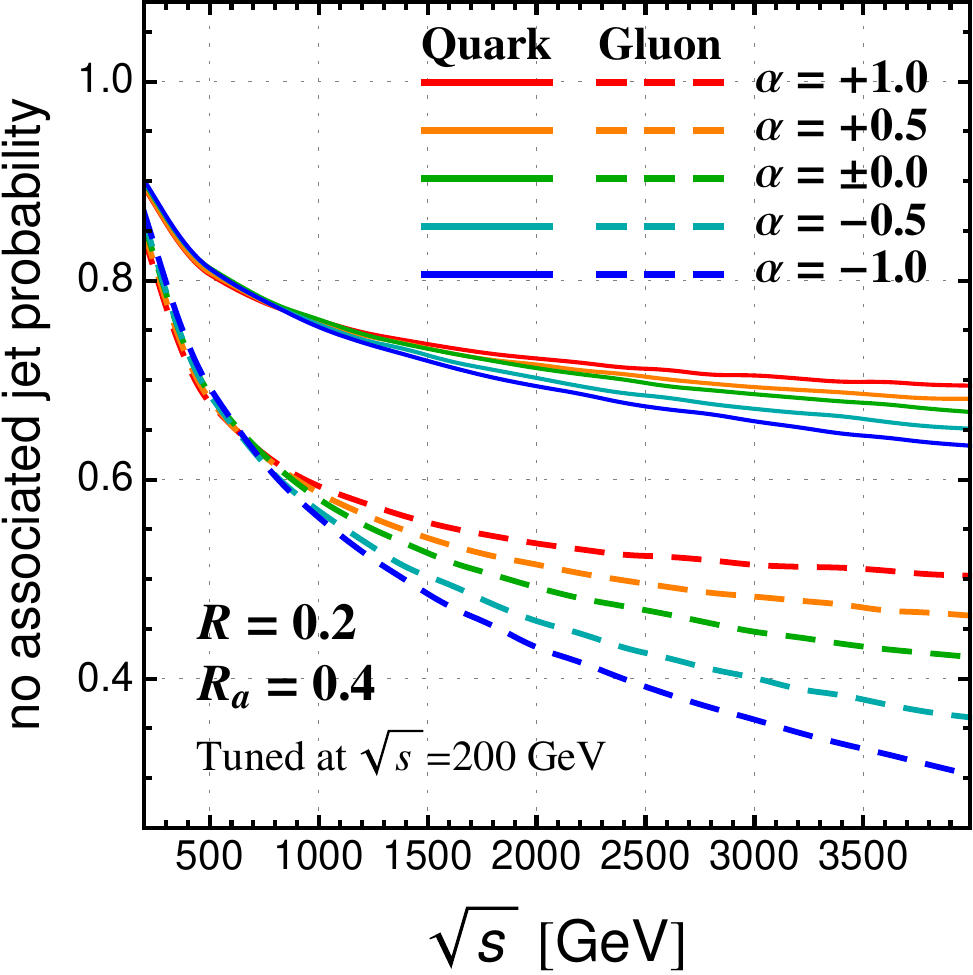}~~~~~~
\includegraphics[width=6.7cm, bb=0 0 280 280]{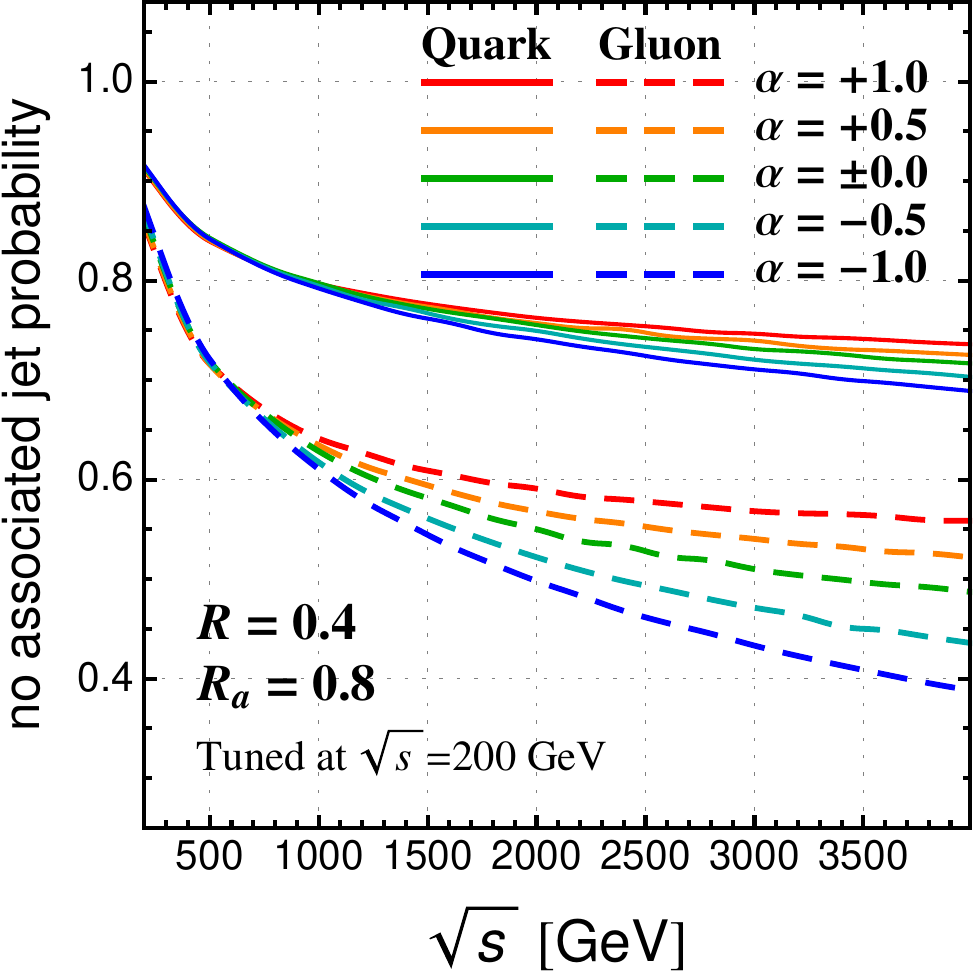}
\caption{{\footnotesize 
No associated jet probabilities for $(R,R_a)=(0.2,0.4)$ and $(0.4,0.8)$, computed with the input parameters as in Table \ref{FittedPara}. 
}}
\label{Ajet}
\end{center}
%
\begin{center}
\includegraphics[width=6.7cm, bb=0 0 280 280]{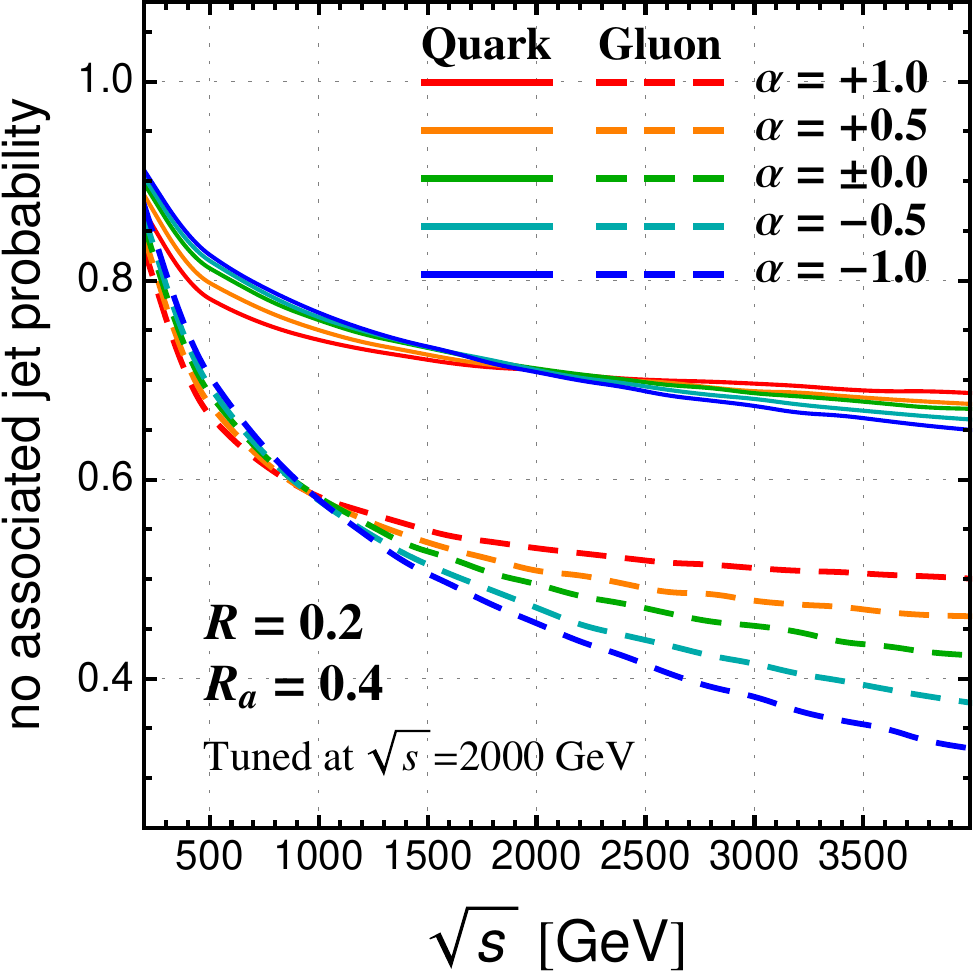}~~~~~~
\includegraphics[width=6.7cm, bb=0 0 280 280]{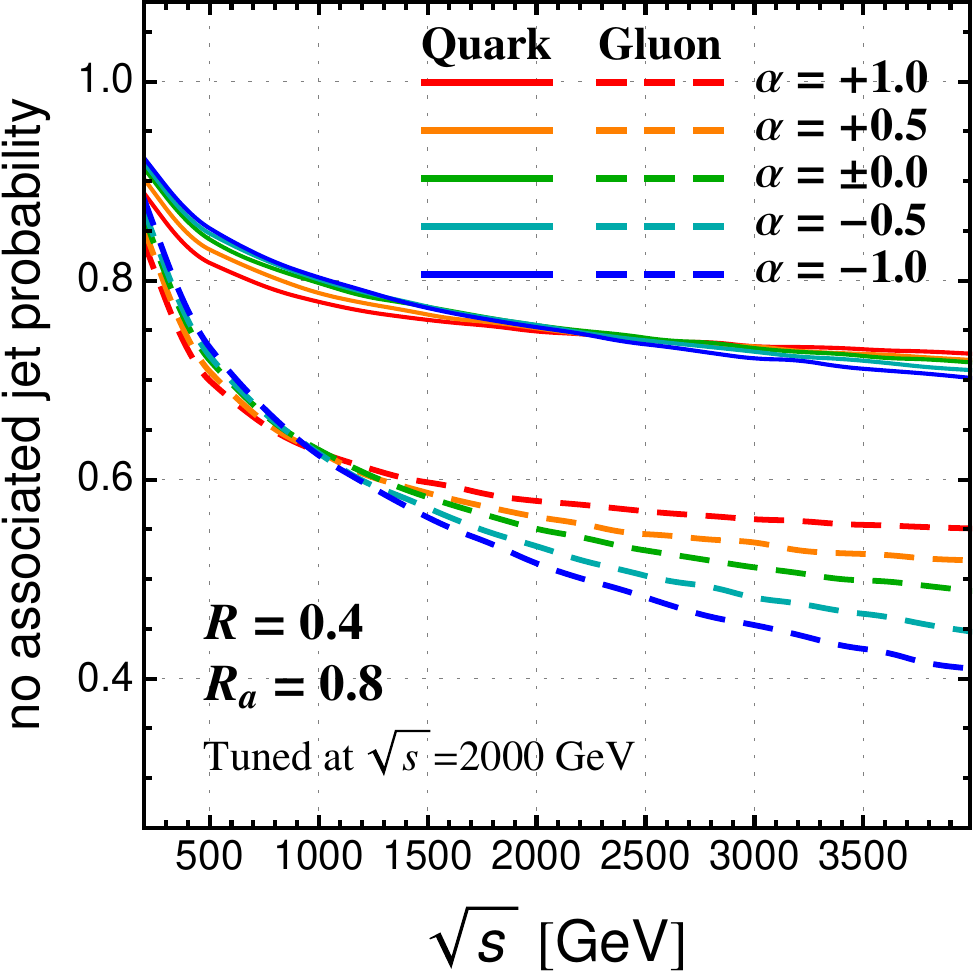}
\caption{{\footnotesize 
Same as Fig. \ref{Ajet}, with the input parameters obtained by fitting the $\ln C_1^{(0.5)}$, $\ln C_1^{(2.0)}$ and $\ln C_1^{(3.0)}$ distributions for quark jets in $e^{+}e^{-} \to q\bar{q}$ events at $\sqrt{s}=2000$ GeV.}}
\label{Ajet_Q2000}
\end{center}
\end{figure}

The no associated jet probabilities calculated with $Q_{1}$, $Q_{0.5}$, $Q_{0}$, $Q_{-0.5}$ and $Q_{-1}$ are shown in Fig. \ref{Ajet}, where, the fitted values of the parameters in Table \ref{FittedPara} have been used. We can see that no associated jet probabilities are similar for each $\alpha$ at the low energy range. This is expected as the parameters have been tuned at $\sqrt{s}=200~{\rm GeV}$. The $\alpha$ dependence is enhanced at the high energy range. 
The larger $\alpha$ is, the larger the no associated jet probabilities become. Therefore, an angular ordered shower $(\alpha =-1)$ predicts wider jets, while a $p_{\bot}$ ordered shower $(\alpha =1)$ predicts narrower jets. This result is qualitatively in agreement with the missing phase space of the $p_{\bot}$ ordered shower \cite{Webber:2010vz}. The wideness of the emissions in the jets are thus tunable by changing the parameter $\alpha$ in the evolution variable continuously.

Fig. \ref{Ajet_Q2000} is similar to Fig. \ref{Ajet}, with the tuning parameters obtained by fitting $\ln C_1^{(\beta)}$ distributions for quark jets in $e^{+}e^{-} \to q\bar{q}$ events at $\sqrt{s}=2000$ GeV. The no associated jet probabilities are similar for each $\alpha$ around $\sqrt{s}=2000$ GeV for the quark jets.  
The $\alpha$ dependence now appears at other energy ranges. The energy scaling of the wideness seems to be inherent in the choice of the evolution variable for the same modelling of the parton shower.

\section{Conclusions}
\label{Conclusions}
In this paper, we have introduced a generalized evolution variable $Q_{\alpha}$ which is a function of the free parameter $\alpha$ taking continuous values.
Although the evolution equation governing the QCD radiation in jets takes an equivalent form for each $\alpha$,  jet substructure depends on $\alpha$ even in the same parton shower formalism. We have examined the $\alpha$-dependence of $C_1^{(\beta)}$ distributions and the associated jet probability for quark and gluon jets. This is motivated by the differences found in the prediction for jet substructure observables between often-used Monte Carlo generators, and 
also by the fact that recent LHC data related to QCD jet substructure lies between the predictions of the MC generators. 
The angular-ordered parton shower formalism used in this study is built upon the one implemented in {\tt Herwig++}. We leave further studies based on other recent parton shower formalisms to a future work.

We have studied the distributions of the first, second and third emissions in the momentum fraction $z$ and emission angle $\theta$ plane. These distributions are of importance as the beginning emissions in the jets have a significant impact on $C_1^{(\beta)}$  and other jet shape observables. The distributions show a unique emission pattern for each choice of $\alpha$. 

We have tuned the parameters in the parton shower to  $e^+e^- \to q\bar{q}$ mock data generated using {\tt Herwig++}, with center of mass energies of $\sqrt{s}=200$ GeV and $2000$ GeV. Observables used in the tuning are $\ln C_1^{(0.5)}$, $\ln C_1^{(2.0)}$ and $\ln C_1^{(3.0)}$ distributions with the jet cone angle $R=0.4$. From this fit, we observe that larger values of the strong coupling are preferred as we vary the values of $\alpha$ from $-1$ to $1$. 
This is qualitatively in agreement with previous findings regarding the difference between the parton shower phase-space covered by the $p_\bot$ ordered and angular ordered showering algorithms. Using the best fit parameters, we have calculated the $\ln C_1^{(\beta)}$ distributions of the quark and gluon jets, with $R=0.4$ and $1.2$, for $e^+e^-$ collisions at $\sqrt{s}=200$ and $1000$ GeV. As we move away from the setup used for the fits (namely, quark jets, $R=0.4$, $\sqrt{s}=200$ GeV),  the $\alpha$-dependence becomes more apparent, especially for larger values of $\beta$ in $C_1^{(\beta)}$. 

The $\alpha$-dependence for large $\beta$ implies that wideness of the soft emissions, especially the first ones in a jet are different for each $\alpha$. 
We can examine this wideness directly by studying the associated jet probability. A high probability for having no associated jet simply means that the probability of wide emissions occurring around a hard jet is low. We have found that the larger $\alpha$ is, the larger the no associated jet probability becomes. This gives us a qualitative understanding of the generator dependence of associated jet rates, especially between {\tt Pythia8} and {\tt Herwig++}. Our results open up the possibility that we might be able to reproduce the wideness of jets observed in real data by varying the value of $\alpha$ in the evolution variable continuously.

\section*{Acknowledgements}
I thank Satyanarayan Mukhopadhyay for helpful discussions and his careful reading of the manuscript, and also Mihoko M. Nojiri and Bryan R. Webber for useful comments. This work is supported by JSPS KAKENHI No.~15J06917.



\end{document}